\documentstyle[aps]{revtex}
\begin{document}
\twocolumn
\title{Non-equilibrium Thermodynamics approach to\\
Transport Processes in Gas Mixtures}
\author{Z. Hens and X. de Hemptinne}
\address{Department of Chemistry, Catholic University of Leuven,\\
Celestijnenlaan 200 F, B-3001 Heverlee, Belgium}
\maketitle

\begin{abstract}
The thermodynamic approach to non-equilibrium dynamics describes the state
of macroscopic systems by means of a collection of intensities or intensive
variables. The latter are by definition the differentials of the entropy with
respect to the set of extensive constraints. The environment is directly
involved in controlling the intensities. The isolation paradigm is negated. The
general principles substanciating the approach are restated and expanded to
multi-component systems. The procedure is applied to the prediction of
transport processes (viscosity and thermal conductivity) of mixtures of atomic
gases. Theoretical results are compared with published experimental data.
\end{abstract}


\section{Introduction}
Excepting recent work centred on trajectory calculations \cite{gasp:92,gasp:95},
the common parenthood of most theoretical approaches to the evolution of
irreversible phenomena traces back to Boltzmann's dynamic equation
\cite{lebA:93,lebB:93}. The core of the model is based on the isolation
paradigm.
Interaction with the surroundings is claimed not to be the necessary condition
for breaking the correlations representing particular non-equilibrium initial
conditions. By contrast, the molecular chaos hypothesis is presumed. Although
this has been violently criticized in the early days, there is presently a
considerable literature in mathematical physics arguing in favour of this
initial
assumption \cite{lebA:93,sinai:76,cornfeld:82}. Some claim that chaotic
Hamiltonian dynamics is the key to decorrelation leading to the required
molecular chaos conditions \cite{prig:88}.

Contrasting with the former, some other schools of thoughts suggest that
correlation breaking is a property typically imported from the surroundings
\cite{posch:88,evans:90,salmon:82}. They confirm that strictly isolated systems
would follow Liouvillian dynamics with conservation of entropy. The apparent
success of Boltzmann's equation and its different approximations is claimed to
result from its {\it a priori} introduction of the molecular chaos assumption.
The latter masks or hides non-Hamiltonian dissipative components equally
present in the global dynamics, as the effect of unavoidable interactions with
the external world \cite{xh:85,xh:92}.

One of the arguments of the followers of the isolation paradigm is that
transport properties characterizing fluid systems and typical of irreversible
dynamics (viscosity, thermal conduction etc.) are bulk properties, whereas
interaction with the environment is a surface phenomenon. To this assertion
let it be opposed that viscous flow implies the presence of walls, coupled to
the system, where the relevant collective momentum is input or withdrawn.
Furthermore, referring to the celebrated Joule experiment, where compressed
gas is made to expand spontaneously in an evacuated vessel, it is easy to
demonstrate that the initial acoustic shock requires intervention of the walls
in order to reach final relaxation. The softer the walls, the faster does the
acoustic perturbation vanish. In the unphysical extreme hypothesis where no
coupling at all to the environment would exist, not even that represented by
action of the ubiquitous electromagnetic radiation field (black body), the
system's dynamics would be governed by its only conservative Hamiltonian.
The motion would then retain at all times the memory of its initial conditions.

Considering that the dynamics of irreversible processes implies necessarily at
some stage action of the surroundings, a general procedure has been
developed that quantifies the latter's most relevant properties. This is
conveniently done by referring to thermodynamics, to be extended to
conditions out of equilibrium. Thermodynamics opens indeed the door to the
definition of intensive variables conjugate to every extensive property. For
given properties, differences of the intensities in and out the system measure
how much the system and its environment are removed from their mutual
equilibrium conditions. The procedure has been elaborated and discussed
extensively elsewhere \cite{xh:85,xh:92,xh:95,xh:96}. It has been applied
successfully to a manifold of simple relaxing systems and to systems
supporting steady transport of extensive properties (energy, momentum etc.).
It has been shown to remain valid in conditions far removed from equilibrium,
even beyond predicted bifurcations. The present work expands some of the
early results to mixtures of gases and compares the theoretical expectations
to published experimental data.

In the traditional approach by Boltzmann and his followers, unbound free flow
implies the usage of a Lagrangian description to formalize the motion
\cite{ryhm:85}. That is why, using symbol $f$ to represent the distribution
function in phase space, and omitting extraneous forces for simplicity,
Boltzmann's dynamic equation is written \cite{evans:90}
\begin{equation}                               
{{\partial f}\over{\partial t}}+ \sum_k v_k {{\partial f}\over{\partial x_k}}
=C(f),
\end{equation}
where $C$ represents the collision integral. The resulting distribution
functions are time dependent, even in stationary conditions \cite{hirsch:54},
leading to physically less transparent conclusions. By contrast, by referring
relaxing systems to their fixed boundaries and considering local thermodynamic
properties, the thermodynamic approach fits remarkably in an Eulerian frame
\cite{ryhm:85,goldstein:51}
\begin{equation}                               
{{df}\over{dt}}=[f,H]+J.
\end{equation}
In the latter equation, $[f,H]$ is the Poisson bracket describing the implicit
motion while $J$ is a source/sink term expressing explicit action of the
environment. Integrating the equation yields very simple expressions for the
relevant distribution functions. In stationary conditions the distribution
functions are time independent. This a great advantage of the procedure.

The present contribution is structured as follows. General principles
supporting thermodynamics of systems out of equilibrium are outlined in
section 2. The formalism leading to prediction of transport properties is
developed in section 3, where it is applied to single component dilute gases.
Extension towards mixtures of the atomic gases and comparison with published
experimental data is the subject of the last section.

\section{Thermodynamics out of equilibrium}
It is impossible to specify exactly the state of a complex macroscopic system
(macrostate). We must content ourselves with descriptions that are
considerably less than complete. In fact, our exact information about the
properties of many-particle systems is restricted to a small number of
mechanical observables or constraints, directly related to the system's
Hamiltonian. Among others, let us quote: the number of particles of any sort
($j$): $N_j$, the total energy $E$ and the accessible physical volume $V$. That
are the traditional micro-canonical constraints or extensive variables. In non-
equilibrium conditions, additional constraints prevail, like the total linear
momentum ${\pmb P}$, the total angular momentum and also possible momenta
of all the properties cited above.

\subsection{The Entropy}
Any function determined completely by the set of constraints describing its
particular macrostate is a function of state. The theoretical definition of the
function of state {\it entropy} goes back to Boltzmann.
\begin{equation}                                 
S=k_B \ln [W(A)].
\end{equation}

For the inventor, $W(A)$ meant ``wahrscheinlichkeit'' which is probability.
Digging for the realities hidden behind this word may lead to some
controversies but, using the same initial letter, most authors wisely prefer the
English {\it weight of the given observational state or macrostate}. The latter
is interpreted as the measure of the domain accessible to the motion in phase
space, given the set of constraints (represented here by the collective
variable $A$) describing the system's particular macrostate. An equivalent
definition for $W(A)$ is the number of quantum states (= microstate) all
compatible with the given set of constraints. Let it be stressed that
Boltzmann's definition is applicable to non-equilibrium conditions simply by
including the additional constraints in the definition.

Let the list of the extensive constraints defining a given macroscopic system
in a particular macrostate be written $\{X_r\}$. The entropy is a function of
this collection of variables. By differentiating with respect to this set we get
by definition the set of conjugate intensive variables or intensities
$\{\xi_r\}$.

\begin{equation}                                     
d S= \sum _r\, {{\partial S}\over{\partial X_r}}\,d X_r=-k_B\,
\sum_r \,\xi_r\,dX_r.
\end{equation}

Equation (4) is Gibbs' celebrated differential equation, generalized
to possible non-equilibrium conditions. It renders the usual temperature
$(\partial S/\partial E)^{-1}$ and the collection of chemical potentials
$-T(\partial S/\partial N_j)$. In non-equilibrium conditions it generalizes the
definition by attaching a given intensity to each of the additional
non-equilibrium constraints.

Boltzmann's definition of the entropy is valid whatever the number of particles
in the system of interest. A significant advantage of referring to its
differentials, namely the intensities, is that their values are independent of
this number and also of the discrete nature of physical systems.

\subsection{Generalized Massieu function}
If two systems are allowed to exchange some extensive properties it is easy
to show that the state of mutual equilibrium, that is the condition where
exchange vanishes on the average, occurs when the conjugate intensities
equalize \cite{kittel:61}. The total entropy becomes then insensitive to
possible
infinitesimal fluctuations in the relevant exchange \cite{xh:92}.

Let the two systems to be considered be a huge body representing the
surroundings (reservoir) on one hand and a small object called the system on
the other. Their respective dimensions are such that thermodynamic flows do
not alter significantly the reservoir's intensive variables. The latter are
therefore constants, defining the external experimental parameters and fixing
the constraints imposed to the smaller system. Intensities are indeed better
measured and controlled. Therefore, instead of referring to the entropy, an
explicit function of the extensive properties ($\{X_r\}$), thermodynamics makes
widely use of thermodynamic potentials and Massieu-Planck functions, obtained
from the extensive properties by Legendre transformations \cite{lands:61}.

The constant volume Legendre transform of the entropy is the generalized
Massieu function ${\cal M}(V,\xi_r)$. It is obtained by performing the following
transformation on the entropy:
\begin{equation}                         
{\cal M}(V,\xi_r)={S \over{k_B}}+\sum_r \xi_r X_r,
\end{equation}
whereby the volume is not included in the collection of indexed constraints.

Differentiating ${\cal M}$ with respect to any intensive variable yields readily
the conjugate extensive properties.
\begin{equation}                         
X_r={{\partial {\cal M}}\over{\partial \xi_r}}.
\end{equation}

\subsection{Dilute Gases}
With ideal gases, the expression for the generalized Massieu function takes a
very simple form. Individual particles assumed to be independent, the global
motion may be represented by a swarm of points in a reduced
$6N$-dimensional single-particle phase space ($\Gamma$).

Let $f(\Gamma)$ be a particle distribution function. Any extensive property
$X_r$ may then be related to a generating function $\phi_r (\Gamma)$ so that
\begin{equation}                                  
X_r=\int_{\Gamma} \phi_r(\Gamma) f(\Gamma) d \Gamma.
\end{equation}
With that formalism, using the Lagrangian multipliers procedure to specify the
maximum entropy conditions compatible with the set of constraints, it is easy
to show \cite{xh:92} that function $f(\Gamma)$ becomes
\begin{equation}                                  
f(\Gamma)= \exp [\sum _r \xi_r \phi_r(\Gamma)].
\end{equation}

With ideal gases, if equation (5) is implemented with the two latter
expressions,  ${\cal M}$ takes the very simple form
\begin{equation}                                  
{\cal M}(\xi_r,V)= \int_{\Gamma} \exp [ \sum_r\xi_r \phi_r(\Gamma)]\,d \Gamma.
\end{equation}
The numerical value of this function is the (average) number of particles
contained in the system. Through the integration limits in configuration space
it has the system's physical dimensions (volume) as one of its independent
variables. By restricting the integration to the only momentum coordinates, a
local generalized Massieu function is obtained, the value of which represents
the average local density in configuration space.

With real gases, the generalized Massieu function is modified due to the
interaction potential between the particles. The simplified formulation is
however still useful as an approximation in low density conditions, when the
duration of the inter-particle collisions is negligible compared to the time
separating collisions. With hard spheres this is certainly the case.

\section{Transport coefficients}
One of the main objectives of the theoretical approach to non-equilibrium
dynamics is the prediction of transport coefficients from first principles.
Comparison between the predicted and experimental results is often considered
as a test for the validity of the relevant attempt.

Since Boltzmann first proposed his kinetic equation there has been a
considerable literature concerning the calculation of the transport coefficients
\cite{balescu:75,clarke:76}. Most frequently cited are the traditional Chapman
and Enskog derivations \cite{hirsch:54} and the Green-Kubo formalism.

It has been stressed above that, for all but perhaps a few mechanical
properties, exchange occurs more or less readily with the surroundings,
tending to equalize the conjugate intensities to the reservoir values. This
justifies thermodynamic expressions based on intensities. When intensities
conjugate to exchangeable properties are different from the reservoir values,
we have transient conditions from where the system tends to relax. If the
system of interest is connected to a surroundings that is not at equilibrium,
it reaches and remains in a stationary state out of equilibrium. This is the
condition we shall focus on now.

If the system is interacting with two reservoirs at different temperatures
separated by some distance (here: $2D$), the conditions of the surroundings
define and dictate to the system the genuine non-equilibrium intensity
``temperature gradient''. Similarly in the Couette flow problem, the externally
imposed gradient is caused by a couple of walls moving in opposite directions.
This generates in the system a non-equilibrium intensity ``gradient of shear
momentum''. Asymmetric exchange with the two reservoirs produces flows. In
this section, the relevant transport coefficients will be examined using a
thermodynamic description. For simplicity, the discussion will however be
limited to hard sphere atomic gases.

In extremely low density systems, where the mean free path is comparable or
longer than the system's physical dimensions (Knudsen gas), properties picked
up by any particle from one wall are transported in a single jump to the
opposite wall. Transport is very efficient indeed. In the thermodynamic limit
(non-Knudsen regime), head-on collisions of like particles do not slow down
the transport properties. By contrast, parallactic or off-axis inter-particle
collisions do. Their effect is one of reducing the range of free transport,
while information about the conditions prevailing in the external reservoir and
available at the boundaries is transferred to the relevant region of the bulk.
As a result, local values of the thermodynamic properties are justified.

For the same reason, the flow rates depend on the average periodicity $\tau$
of the perturbing collisions.

\subsection{Single component gases}
Let us consider an arbitrary property $X_r$ with generating function
$\phi_r(\Gamma)$. We assume that its flow is directed along the {\it z}-axis.
Let us consider a plane positioned at coordinate $z^{*}$. The basic equation
for the flow $J_r$ of the relevant property through this plane is
\begin{eqnarray}                                          
\nonumber J_r = {1 \over \tau}\;\int\!\int\!\int {{d^3{\pmb p}}\over h^3}
\int_{(z^{*}-p_z \tau/m)}^{z^*} \qquad \qquad \qquad \\
\nonumber \qquad \\ \qquad \qquad
\phi_r(\Gamma)\exp[ \sum_l\xi_l \phi_l(\Gamma)]\,dz.
\end{eqnarray}
The symbol $d^3{\pmb p}$ is a short form for $dp_x dp_y dp_z$. Integration
limits in momentum space is $-\infty \rightarrow \infty$. Plank's constant in
the numerator represents a norm in phase space, related to the classic-
quantum correspondence \cite{kittel:61}. In the subsections to follow, this
equation will be applied to different types of flow.

It may be useful to stress the difference between equation (10) and that
proposed elsewhere, in own relevant different contexts for flows of extensive
properties \cite{hirsch:54,clarke:76}:
\begin{equation}                                     
J_r = \int\!\int\!\int{{p_z}\over m}\,\phi_r(\Gamma)\,\exp[ \sum _l\xi_l
\phi_l(\Gamma)]\,{{d^3{\pmb p}}\over h^3}.
\end{equation}
Contrasting with the latter, equation (10) confirms that transport occurs
during the free motion time separating relaxing collisions. The lesser the
collision frequency, the more effective is the transport. Collisions do not
activate transport. They do increase resistance to transport.

\subsubsection{Viscosity}
We consider a fluid bound by a pair of walls moving in opposite directions
(Couette flow). Excepting negligible higher order corrections (proportional to
$\tau^2(\beta m D^2)^{-1}$), the system's conditions are defined completely (at
the lowest order in $\tau$) by the set of constraints listed in table
\ref{viscosity}. The intensities under direct control of the surroundings
(exchangeable) are the particles number, the kinetic energy and the intensity
conjugate to the gradient of shear momentum. It may indeed be verified
readily that the velocity of the walls ({\it y- }direction) equals $\pm
\sigma_y/\beta$. Two variables remain to be determined, namely $\theta_2$ and
$\gamma_2$. (Index 2 refers to quadratic moments). They require two
independent equations.

In stationary conditions there is no local accumulation of the transverse
component of momentum ($p_z$) (no pressure gradient, no acoustic
perturbation). The relevant flow is therefore independent of $z^*$. Likewise,
the total flow of energy through the system is zero. By implementing equation
(10) with the two relevant generating functions, the conditions $\partial
J_{p_z} / \partial z =0$ and $J_{U}=0$ yield together
\begin{equation}                                     
\theta_2=0,\qquad \qquad
{5 \over 2} \;{{\gamma_2}\over\beta}={{m\sigma^2}\over{2\beta}}.
\end{equation}

The flow of shear momentum may now be determined by implementing equation
(10) with the generating function $p_y$, where $\theta_2$ and $\gamma_2$
have been replaced by their values. This yields
\begin{equation}                                     
J_{p_y}=-{{\sigma_y}\over{2\beta D}}\;{{n \tau} \over{\beta }},
\end{equation}
where $n={\cal M}/V$ represents the particle density.

The phenomenological reaction at the plates' level compensating for transfer
of momentum from wall to wall is friction. Shear viscosity is the ratio of the
sum of the forces applied to the two plates to the velocity gradient
($\sigma_y\,(\beta D)^{-1}$). Following equation (13), its value is
\begin{equation}                                     
\eta=n\; {\tau\over\beta}.
\end{equation}

\subsubsection{Thermal conduction}
Now we consider a system in thermal contact with a couple of heat reservoirs
at different temperatures separated by a distance $2D$. The system's
stationary non-equilibrium conditions are completely described (at the lowest
order in $\tau$) by the set of constraints listed in table \ref{conduction}. By
inspecting the generating function conjugate to the temperature gradient it
is clear that $k_B\nabla T=-\gamma_1 \, (\beta^2D)^{-1}$. (Index 1 refers to
linear moments).

Knowing that the system is bound by a pair of impervious walls it may look
strange that collective motion of the particles perpendicularly to the walls
needs to be anticipated in constructing the expression for the non-equilibrium
distribution function $f$ (equation 8).

In a system controlled by two heat reservoirs at different temperatures,
kinetic energy is not uniformly distributed among the particles. Those moving
towards the cold wall have been equilibrated with the system upstream in a
hotter region at the instant of their previous collision and vice-versa. In
moving from the hot wall to the cold one, particles travel on the average
faster than in their return cycle. If the particles are to change their average
kinetic energy in a correlated fashion on impact with either walls, while the
container (the pair of walls) is to remain immobile, collective momentum is
transferred by the container into the system.

The intensities under direct control of the surroundings (conjugate to
exchangeable properties or otherwise given constant properties) are the
intensity conjugate to the particle number, the temperature (or better $\beta$)
and the temperature gradient. Two intensities remain to be determined:
$\theta_1$ and $\sigma_z$. This requires two independent equations. One is
the condition for stationarity. The other equation describes mechanical
equilibrium of the system between its walls.

According to the local description, at any point in the bulk of the system, the
average particles density $n$ responds to the equation
\begin{eqnarray}                                          
\nonumber n(z^*)={1\over h^3}\int_{-\infty}^{\infty} dp_x \int_{-
\infty}^{\infty} dp_y \int_{-\infty}^{\infty} dp_z \qquad \qquad \\
\nonumber \qquad \\ \qquad \qquad
\exp \{ \sum _r \xi_ r \,\phi_r [(z=z^*),{\pmb p}]\}.
\end{eqnarray}

At any position $z^*$, we consider the partial density $n_+(z^*)$ of the only
particles with positive velocity along the {\it z}-direction. Stationarity
implies
that this partial density is compensated exactly by the sum of the densities
of the particles issued from regions from where they will be reaching this
position undisturbed in one collision period, their velocities being oppositely
oriented ($p \leq 0$). Hence
\begin{eqnarray}                                          
\nonumber n_-(z^*)={1\over h^3}\int_{-\infty}^{\infty} dp_x \int_{-
\infty}^{\infty} dp_y \int_{-\infty}^0 dp_z \qquad \qquad \\
\nonumber \qquad \\ \qquad
\exp\{ \sum_r \xi_r \, \phi_r [(z=z^* - {{p_z \tau}\over m}),{\pmb p}]\}.
\end{eqnarray}
To the lowest order in $\tau$, Relation $n_-(z^*)=n_+(z^*)$ yields
\begin{equation}                                          
\bigl(\theta_1-2 {{\gamma_1}\over\beta}\bigr)\;{{\tau}\over{mD}}=2\sigma_z.
\end{equation}

The second equation expresses position independence of flow of momentum
across the system. In other words, there are no pressure gradients. Equation
(10) is used with $p_z$ as the flow defining generating function. The condition
$\partial J_{p_z}/\partial z=0$ yields (to the lowest order in $\tau$)
\begin{equation}                                         
\theta_1={5 \over 2}{\gamma_1 \over \beta}.
\end{equation}
Flow of energy (heat) through the system is given by this same general
equation (10) where the flow defining generating function is now
$\sum(p^2/2m)$. For particles associated with internal rotational motion
(Eucken correction \cite{hirsch:54,eucken:13}), the relevant contribution to the
energy should be added to the latter generating function. With atomic gases
the result reads
\begin{equation}                                     
J_E={5 \over 2}{n \over {\beta^2}}\bigl[\sigma_z - {1 \over 2}\bigl(
\theta_1-{7 \over 2}{{\gamma_1}\over \beta}\bigr){\tau \over {mD}}\bigr].
\end{equation}
By implementing the latter with the relevant values of $\theta_1$ and
$\sigma_z$, flow of energy becomes
\begin{equation}                                          
J_E={{15}\over 8}\;{{\gamma_1}\over{\beta^2D}}\;n\;{\tau \over {\beta m}}.
\end{equation}

Heat conductivity ($\lambda$) is defined as the ratio between the sum of the
rates of heat exchange at either walls ($2J_E$) and the temperature gradient.
Hence
\begin{equation}                                          
\lambda={{15}\over 4}\;k_B\;n\;{\tau\over{\beta m}}.
\end{equation}

In equations (14) and (21) the transport coefficients are expressed in terms
of the effective collision periodicity $\tau$. For direct comparison with
experimental results, an additional expression is required that relates the
collision periodicity to the mechanical properties of the colliding species
(mass
and cross-section) at the given temperature. Here there remains an
uncertainty concerning the model to be adopted for relaxing collisions,
yielding the effective cross-section as a function of the temperature. For
evaluating the results presented above it is therefore advisable to eliminate
the variable $\tau$ in the discussion. That is where the Prandtl number comes
in. It is related to the ratio between viscosity and heat conductivity and
given by
\begin{equation}                                          
Pr={{\eta \, c_p}\over{m\,\lambda}},
\end{equation}
with $c_p$ as the constant pressure heat capacity. By implementing this
definition with the results obtained above, the experimental values are
obtained identically, thereby corroborating the general model.
\cite{hirsch:54,clarke:76}.

\subsection{Mixtures of atomic gases}
We are now investigating transport processes in binary mixtures of gases. Let
the components be indexed $A$ and $B$, where $A$ points to the component
with the highest mass. Each may be considered as a separate system, with its
own thermodynamic functions, interacting simultaneously with the other
component and with the environment. The generalized Massieu function being
an extensive properties, we have for the composite system
\begin{equation}                               
{\cal M}={\cal M}_A + {\cal M}_B.
\end{equation}

With dilute gases or gases interacting as hard spheres, the individual
generalized Massieu functions are defined as in equation (9). For each
component separately the generating functions to be used are the same as for
single-component gases (see tables \ref{viscosity} and \ref{conduction}),
excepting for the additional indexing of the mass of the relevant particles in
the generating function for kinetic energy. In stationary or quasi-stationary
conditions (see below), for exchangeable properties where equilibrium between
the subsystems prevails, the intensities are the same. In the examples treated
below, that will be the case for the temperature and its moments and for the
intensities conjugate to the collective motion. Intensities conjugate to the
populations and their own moments however will be indexed according to which
component they refer to.

It has been stressed above that the collision periodicity is an essential
ingredient in the dynamics of systems out of equilibrium. In multi-component
systems, there is an average collision periodicity for each of the constituents
($\tau_A, \tau_B$). It measures for each component how long the relevant
atoms move freely before being halted by the matrix formed by the other
particles, making them feel the thermodynamic conditions dictated by the
environment.

In multi-component systems there are homogeneous and heterogeneous relaxing
collisions. Their frequencies add up. The efficiency for exchange of momentum
from a colliding atom to the local thermodynamic bath depends on the masses
of the collision partners. When a heavy particle hits a light constituent of the
thermodynamic bath, its path is less disturbed and less momentum is
transferred than in the opposite case.

We assume a particle with mass $m_1$ and linear momentum ${\pmb P}$ hitting
a stationary matrix particle with mass $m_2$. If the exit path of the matrix
particle forms an angle $\psi$ with the incident one, momentum transferred to
the matrix equals $2|{\pmb P}|\cos(\psi)m_2/(m_1+m_2)$. Hence, the relative
transfer efficiency of heterogeneous collisions is $2m_2/(m_1+m_2)$. For the
total effective collision frequency of atoms of one sort with respect to the
matrix (the reciprocal of $\tau$), the latter coefficient is the appropriate
scaling factor relating the efficiency of heterogeneous collisions to the
homogeneous ones.

We take the atoms to be hard spheres. The collision cross-sections are
respectively $d_{AA}, d_{BB}, d_{AB}$. Using the scaling parameter defined
above, omitting the common factor ${5 \over{16}}\sqrt{\beta/\pi}$ and adding
for either constituents the individual effective collision frequencies, the two
total effective collision periodicities read
\begin{eqnarray}                         
\tau_A \simeq \Bigl({{n_A d_{AA}^2}\over \sqrt{m_A}} +
{{2m_B}\over{m_A+m_B}}{n_B d_{AB}^2} \sqrt{{m_A+m_B} \over\ {2m_A m_B}}
\Bigr)^{-1}, \\
\tau_B \simeq \Bigl({{2m_A}\over{m_A+m_B}}n_A d_{AB}^2
\sqrt{{m_A+m_B}\over{2m_Am_B}}+{{n_B d_{BB}^2}\over \sqrt{m_B}}\Bigr)^{-1}.
\end{eqnarray}
Considering that transport coefficients of mixtures are systematically compared
or normalized to either one of the pure gas values, the common factor cancels.

For simplifying the formalism it is advisable to replace $n_A$ by $xn$ and
$n_B$ by $(1-x)n$. This will be done systematically below.

With hard spheres we have $d_{AB}=(d_{AA}+d_{BB})/2$. It appears that
experimental accuracy of the published data on the viscosity of mixtures of
atomic gases is sufficient to allow the heterogeneous hard sphere diameter to
be corrected by a factor $\epsilon$ close to 1.

\subsubsection{Viscosity}
We consider a binary mixture of atomic gases bound by a pair of walls moving
in opposite directions (Couette flow). The mole fraction of substance $A$ is
written $x$ ($N_A+N_B=N$).

The intensities under direct control of the surroundings (exchangeable
properties or otherwise independent properties) are the intensities conjugate
to the particle numbers of either substances ($\alpha_A,\, \alpha_B$), the
temperature (or better $\beta$) and the linear moment of shear velocity
$\sigma_y$ (see table \ref{viscosity}). Three intensities need still to be
determined, namely the quadratic moment of the temperature (or better
$\gamma_2$) and the quadratic moments of the particle distributions for $A$
and $B$ ($\theta_{2,A}, \,\theta_{2,B}$).

The three additional relations required for completing the thermodynamic
description of the system are of the same vein as those used for Couette flow
in single component gases (see above). For symmetry reasons, flow of shear
momentum is independent of the particular values of the additional intensities.
The principles involved in their determination will therefore be postponed
until the section concerning thermal conductivity and diffusion.

Flow of momentum is supported by either components. For each, the
contribution is given according to equation (10), where the generating
function to be implemented as $\phi_r$ is $p_y$. Integration yields
\begin{equation}                         
J_{p_y}=-n{{\sigma_y}\over{2\beta^2 D}}\;[x \tau_A + (1-x) \tau_B].
\end{equation}
The viscosity of the mixture is therefore
\begin{equation}                         
\eta_{mix}={n \over\beta}\;[x \tau_A + (1-x) \tau_B],
\end{equation}
where $n/\beta$ is the total pressure (${\cal P}$).

In figure \ref{vishexe} the result of equation (27), is plotted for a  mixture
of Xe in He. The experimental results at 291 K published by E. Thornton and
coworkers \cite{toulouvis:75} are indicated on the same graph (experimental
uncertainties $\sim \pm 1\%$). The correction factor $\epsilon$ for
heterogeneous collisions may be estimated by fitting the curve to the
experimental results. The curve obtained without the correction factor
($\epsilon = 1$) is displayed as a dotted curve.

The same fit has been performed on the ten different mixtures of atomic gases
at the same temperature of 291 K. Table \ref{fitviscosity} lists the values of
$\epsilon$ giving the best result for each mixture.

\subsubsection{Diffusion and thermal conduction}
We consider now a binary mixture of atomic gases in thermal contact with a
couple of heat reservoirs at different temperatures separated by a distance
$2D$. The mole fraction of the heaviest substance ($A$) is written $x$.

For each of the two components, the stationary non-equilibrium conditions are
completely described by the set of constraints listed in table \ref{conduction}.
The intensities must be indexed accordingly.

The intensities conjugate to the particle numbers of either substances
($\alpha_A,\, \alpha_B$), the temperature (or better $\beta$) and its gradient
(or better $\gamma_1$) are under direct control of the surroundings. Thermal
interaction between the subsystems removes the necessity of indexing the
latter two intensities.

Three intensities remain to be determined namely the two gradients of the
particle distributions ($\theta_{1,A}, \,\theta_{1,B}$) and the intensity
conjugate to the collective momentum from wall to wall ($\sigma_z$). Hence,
three additional conditions or equations are needed in order to describe the
system completely.

Two of the additional conditions are identical to those discussed for single
component systems. It are mechanical equilibrium and stationarity of the total
particle distribution.

Mechanical equilibrium of the system between its walls implies vanishing total
pressure gradient. It does not require {\it per se} vanishing pressure
gradient for either substances separately. A possible pressure gradient of $A$
is neutralized by an opposite gradient for $B$. The condition is formalized by
stating that the sum of the contributions of either substances to flow of
momentum between the boundaries is position independent
\begin{equation}                         
{{\partial J_{p_z,A}\over{\partial z}}} +
{{\partial J_{p_z,B}\over{\partial z}}} = 0.
\end{equation}
Applying equation (10) with $\phi=p_z$ for each of the two substances leads
to
\begin{equation}                         
x \bigl(\theta_{1,A} - {5 \over 2}{{\gamma_1}\over \beta}\bigr)
+ (1-x) \bigl(\theta_{1,B} - {5 \over 2}{{\gamma_1}\over \beta}\bigr) =0.
\end{equation}
Let it be stressed that the pressure gradient for $A$, is
\begin{equation}                         
\nabla {\cal P}_A={{n_A}\over {\beta D}} \bigl(\theta_{1,A} - {5 \over 2}
{{\gamma_1}\over \beta}\bigr),
\end{equation}
with $n_A=xn$, and {\it mutatis mutandis} for $B$.

The condition for stationarity is defined along the same lines as above
(equations 15--17), where the densities $n_+(z^*)$ and $n_-(z^*)$ are now
understood as the sum of the different components. As a result, the relation
for internal collective motion ($\sigma_z$) becomes (see equation 17)
\begin{eqnarray}                         
\nonumber x\sqrt{m_A}\bigl[\sigma_z - {1 \over 2} \bigl(\theta_{1,A}-2
{{\gamma_1}\over\beta}\bigr)\;{{\tau_A} \over {m_AD}}\bigr] \qquad
\qquad \\
\nonumber \qquad \\
+ (1-x)\sqrt{m_B}\bigl[ \sigma_z - {1 \over 2} \bigl(\theta_{1,B}-2
{{\gamma_1}\over\beta}\bigr)\;{{\tau_B} \over {m_BD}}\bigr] = 0.
\end{eqnarray}

The last condition to be considered concerns mutual diffusion or motion of the
subsystems with respect to each other. By implementing equation (10) with the
generating function $\phi_r=1$ the particle flow of either subsystems is
obtained, according to whether the parameters in the exponential function are
indexed $A$ or $B$. The results are
\begin{equation}                         
J_A={{xn} \over {\beta}}\bigl[\sigma_z - {1 \over 2}\bigl(
\theta_{1,A}-{5 \over 2}{{\gamma_1}\over \beta}\bigr){{\tau_A}\over {m_A D}}
\bigr],
\end{equation}
\begin{equation}                         
J_B={{(1-x)n} \over {\beta}}\bigl[\sigma_z - {1 \over 2}\bigl(
\theta_{1,B}-{5 \over 2}{{\gamma_1}\over \beta}\bigr){{\tau_B}\over {m_B D}}
\bigr].
\end{equation}

The first contribution in either equations (that proportional to $\sigma_z$)
represents collective drag generated in the fluid by correlated effect of the
two walls. This acts on the two subsystems alike. Therefore it does not drive
diffusion of one subsystem with respect to the other one. By contrast,
diffusion is described by the second part of the flow equations. As it may be
verified, this is driven by the relevant partial pressure gradient.

Diffusion coupled to flow of heat and vice-versa are known as the Dufour and
the Soret effects \cite{hirsch:54}. Let us consider substance $A$ as the solute
and $B$ as the solvent. If transport of heat and matter are expressed using
the convenient parameters for the relevant generalized forces (the conjugate
intensities, here $\gamma_1$ and $\theta_{1,A}$), Onsager's phenomenological
equations are retrieved \cite{prig:61,katch:65}.

The diffusive stationary state is reached by differential displacement of the
subsystems with respect to each other. Then we have for either subsystems
vanishing partial pressure gradients. This represents therefore the remaining
constraint for complete thermodynamic description of the stationary non-
equilibrium system. The question is however how fast the diffusive stationary
state may been reached in practical cases when thermal conductivity of multi-
component mixtures is measured.

The experimental procedure for measuring thermal conductivity consists in
preparing an appropriate binary mixture in equilibrium conditions in a
conventional thermostat, the mixture being then introduced between two walls
held or brought at different temperatures. When mechanical equilibration is
established (relaxation of acoustic perturbations), the total pressure is flat
(equation 29). Nevertheless, on transferring the mixture in the region with the
temperature gradient, individual pressure gradients on the subsystems may
have been created, forcing the particles to segregate. If the mixture consists
of particles with different mobility it is expected that establishment of final
stationary conditions is slow. The slower moving particles tend indeed to
remain distributed homogeneously, as they were before the establishment of
the temperature gradient, while the partial pressure gradient of the faster
moving subsystem compensates for the former's resulting partial pressure
unbalance.

When comparing predicted values of thermal conduction to experimental data,
there is uncertainty as to how much the system has been allowed to relax the
slow coupled particle segregation in the relevant measurement. Let us assume
this would not have occurred at all (pseudo-stationary state). The two
subsystems may then be considered as acting independently for all the
properties concerning the particle distributions. They remain however tightly
coupled for all the properties that are promptly interchanged. In particular,
they share the same value of $\beta$ and $\gamma_1$. The intensity
$\sigma_z$ conjugate to the collective momentum generated by the temperature
gradient is also common to the two subsystems. Concerning the latter, its
relation to the other intensities and to the collision periodicities is given by
equation (17). Instead of equation (31) we have now two relations, namely
\begin{eqnarray}                                       
\bigl(\theta_{1,A}-2 {{\gamma_1}\over\beta}\bigr)\;{{\tau_A}\over{m_A D}}
=2\sigma_z, \\
\bigl(\theta_{1,B}-2 {{\gamma_1}\over\beta}\bigr)\;{{\tau_B}\over{m_B D}}
=2\sigma_z.
\end{eqnarray}

By combining the equation for overall mechanical stability (equation 29) with
the two latter ones, an expression for the gradients of the individual partial
pressures may be derived. Writing
\begin{equation}
R={{\tau_A/m_A}\over{\tau_B/m_B}},
\end{equation}
this relation reads
\begin{equation}                                     
\theta_{1,A}-{5 \over 2}{{\gamma_1}\over \beta}={1 \over 2}\,{{(1-R)(1-x)}\over
{(1-x)R +x}}\, {{\gamma_1}\over\beta}.
\end{equation}

In practical cases, when heat conductivity is measured, the system may be
somewhere between the two extreme conditions. The uncertainty concerning
how close diffusion has reached stationarity in the experimental conditions
where the measurements have been performed explains why thermal conduction
data of mixtures are difficult to reproduce. Let us express the uncertainty by
a coefficient $c$ to multiply the right-hand side of equation (37). When
discussing a homogeneous set of data with varying compositions $x$, we
assume for simplicity that the same coefficient is valid.

Transport of heat is supported by either components of the mixture.
\begin{equation}                                     
J_E=J_{E,A}+J_{E,B}.
\end{equation}
For each, the contribution is given according to equation (19), where the
relevant intensities are as determined above. Hence,
\begin{eqnarray}                                     
\nonumber J_E={{xn} \over {\beta^2}}\bigl[\sigma_z - {1 \over 2}\bigl(
\theta_{1,A}-{7 \over 2}{{\gamma_1}\over \beta}\bigr){{\tau_A}\over {m_A D}}
\bigr]  \qquad \qquad \\
\nonumber \qquad \\ \qquad + {{(1-x)n} \over {\beta^2}}\bigl[\sigma_z - {1
\over 2}\bigl(\theta_{1,B}-{7 \over 2} {{\gamma_1}\over
\beta}\bigr){{\tau_B}\over {m_B D}} \bigr].
\end{eqnarray}

In comparing the result with experimental data, coefficient $c$ may be taken
as an adjustable parameter. This exercise has been performed on the data at
291 K published by E. Thornton and coworkers \cite{touloucon:70}. The
coefficients yielding the best fit are listed in table \ref{fitconduction}.
Figure
\ref{conhexe} is an illustration of the results. The accuracy is better than the
announced experimental precision (4\%).

\section{Conclusion}
The theory for transport of extensive properties in dilute mixtures of atomic
gases has been successfully developed in the context of the thermodynamic
approach to non-equilibrium processes \cite{xh:92}. The equations are based
on accurate definitions of intensities conjugate to the set of extensive
properties defining the system's particular macrostate. Intensities are the
natural parameters for external control of macroscopic systems. It may
therefore be claimed that the procedure presented above is characterized by
a greater physical transparancy when compared with the traditional treatments
\cite{hirsch:54}. The number of adjustable parameters it implies is low and
their physical meaning is straightforward.

The theory confirms that inter-particle collisions are not responsible for the
transport processes. By contrast, transport occurs during the free motion time
of the particles in the periods separating collisions. By colliding, the
particles
exchange their mechanical properties with the matrix or bath of the remaining
particles representing the system, the local thermodynamic properties of which
are defined by the external constraints. Permanent interaction of this matrix
with the environment causes correlations between individual motions to
disappear. The effect of collisions is to increase the resistance opposed by the
system to flows.

The theory relies on the hard sphere mutual interaction model. This implies
further the definition of an effective temperature dependent collision diameter
for the different particles involved. The theory is clearly not suitable for
predicting this quantity. It requires trajectory calculations based on the
particular inter-particle interaction potentials. The question is well
documented
elsewhere \cite{hirsch:54}. In the treatment presented above, this problem has
been bypassed by focussing on results obtained at constant temperature.

\begin{figure}                                       
\caption{Predicted viscosity of a mixture of Xe in He, with $\epsilon = 0.98$
(smooth curve) and published experimental data (temperature: 291 K). The
dotted curve is for $\epsilon = 1$.}
\label{vishexe}
\end{figure}
\vskip 1cm

\begin{figure}                                       
\caption{Predicted thermal conductivity of a mixture of Xe in He taking $c=0.5$
(smooth curve) and published experimental data (temperature: 291 K). The
dotted curves are for $c=0$ and $c=1$}
\label{conhexe}
\end{figure}
\vskip 1cm

\begin{table}                                        
\caption{List of the main constraints for Couette flow (distance between the
walls: $2D$)}
\label{viscosity}
\begin{tabular}{ccc}
$X_r$ & $\phi_r(\Gamma)$ & $\xi_r$ \\
\hline
Particles number & 1 & $\alpha$  \\
2nd moment of particle distribution & $[(z/D)^2-1]$ & $\theta_2$ \\
Kinetic energy & $\sum({p^2\over {2m}})$ & $-\beta$ \\
2nd moment of energy distribution & $[({z\over D})^2-1]\;\sum({{p^2}\over
{2m}})$ & $-\gamma_2$ \\
Gradient of shear momentum & $(z/D)\, p_y$ & $\sigma_y$ \\
\end{tabular}
\end{table}
\vskip 1cm

\begin{table}                                        
\caption{List of the main constraints for thermal conductivity (distance
between
the walls: $2D$)}
\label{conduction}
\begin{tabular}{ccc}
$X_r$ & $\phi_r(\Gamma)$ & $\xi_r$ \\
\hline
Particles number & 1 & $\alpha$  \\
Gradient of particle distribution & $(z/D)$ & $\theta_1$ \\
Kinetic energy & $\sum({p^2\over {2m}})$ & $-\beta$ \\
Gradient of energy distribution & $(z/D)\;\sum({{p^2}\over {2m}})$ &
$-\gamma_1$ \\
Collective transverse momentum & $ p_z $ & $\sigma_z$ \\
\end{tabular}
\end{table}
\vskip 1cm

\begin{table}                                        
\caption{Correction $\epsilon$ to the heterogeneous collision diameter of pairs
of gases, obtained by fitting the viscocity of the relevant mixtures to
experimental data (temperature: 291 K)}
\label{fitviscosity}
\begin{tabular}{c|cccc}
 &Ne & Ar & Kr & Xe \\
\hline
He & 1.03 & 1.01 & 1.00 & 0.97 \\
Ne &      & 0.98 & 0.98 & 0.98 \\
Ar &      &      & 1.03 & 1.00 \\
Kr &      &      &      & 0.99 \\
\end{tabular}
\end{table}
\vskip 1cm

\begin{table}                                        
\caption{Correction $c$ to the partial pressure of the pairs of gases, caused
by unrelaxed thermal diffusion, obtained by fitting the predicted termal
conductivity to experimental data (temperature: 291 K)}
\label{fitconduction}
\begin{tabular}{c|cccc}
 &Ne & Ar & Kr & Xe \\
\hline
He & 0 & 0.7 & 0.55 & 0.5 \\
Ne &   & 0.5 & 1 & 1 \\
Ar &      &  & 1 & 1 \\
Kr &      &  &   & 1 \\
\end{tabular}
\end{table}

\end{document}